\documentclass[prc,aps,floatfix,groupedaddress,amsmath,amssymb,twocolumn]{revtex4}

\usepackage{graphicx}
\usepackage{dcolumn}
\usepackage{bm}
\usepackage{color}

\usepackage[section]{placeins}
\usepackage{graphicx,epsfig}
\usepackage{amsmath}
\usepackage{amsfonts}
\usepackage{braket}
\usepackage{fancyhdr}
\usepackage{hyperref}
\usepackage[utf8]{inputenc}
\usepackage[T1]{fontenc}
\usepackage{amsthm}
\usepackage{hhline}
\usepackage{multirow}
\usepackage[figuresleft]{rotating}
\usepackage{gensymb}
\def\beq{\begin{equation}}
\def\eeq{\end{equation}}
\def\beqa{\begin{eqnarray}}
\def\eeqa{\end{eqnarray}}

\begin{document}



\title{Effect of chiral nuclear forces on the neutrino mean free path in hot neutron matter}

\author{Isaac Vida\~na$^1$, Domenico Logoteta$^{2,3}$ and Ignazio Bombaci$^{2,3}$}
\affiliation{$^1$Istituto Nazionale di Fisica Nucleare, Sezione di Catania, Dipartimento di Fisica e Astronomia ``Ettore Majorana'', Universit\`a di Catania, Via Santa Sofia 64, I-95123 Catania, Italy}
\address{$^2$Dipartimento di Fisica ``Enrico Fermi'', Universit\`a di Pisa, Largo B. Pontecorvo 3, I-56127 Pisa, Italy}
\address{$^3$Istituto Nazionale di Fisica Nucleare, Sezione di Pisa, Largo B. Pontecorvo 3, I-56127 Pisa, Italy}


\begin{abstract}
We study the role of chiral nuclear forces on the propagation of neutrinos in hot neutron matter. In particular, we analyze the
convergence of the dynamical structure factor and the neutrino mean free path with the order of the power counting of the chiral forces, as well as the role of the regulator cut-off of these forces in the determination of these quantities. Single-particle energies and chemical potentials needed to calculate the dynamical structure factor are obtained within the Brueckner--Hartree--Fock approximation extended to finite temperature. Our results show that the dynamical structure factor and the neutrino mean free path depend on the cut-off only when the chiral potential is considered at leading order (LO) and next-to leading order (NLO), becoming this dependence strongly reduced at higher orders in the chiral power counting due to the role of three-nucleon forces that start to contribute at next-to-next-to leading order (N$^2$LO) being, in particular, almost negligible at next-to-next-to-next-to leading order (N$^3$LO). The neutrino mean free path is found to converge up to densities slightly below $\sim 0.15$ fm$^{-3}$ when increasing the order of the chiral power counting, although no signal of convergence is found for densities above this value. The uncertainty associated with our order-by-order nuclear many-body calculation of the neutrino mean free path is roughly estimated from the difference between the results obtained at N$^2$LO and N$^3$LO, finding that it varies from about a few centimeters at low densities up to a bit less than $2$ meters at the largest one considered in this work, $0.3$ fm$^{-3}$. 
\end{abstract}


\maketitle


\section{Introduction}
\label{sec:intro}

Neutrinos and, particularly, the knowledge of their interactions in hot and dense baryonic matter  \cite{Tubbs75,Sawyer75,Lamb76,Lamb78,Tubbs78,Bludman78,Sawyer79,Iwamoto82a,Iwamoto82b,Goodwin82,Burrows82,Goodwin82b,Bruenn85,vandenHorn86,Maxwell87,Cooperstein88,Burrows88,Sawyer89,Schinder90,Horowitz91a,Horowitz91b,Horowitz92,Prakash92,Reddy92,Keil95,Janka96a,Sigl96,Reddy97,Reddy98,Reddy99,Navarro99,Reddy00,Prakash01,Margueron03a,Margueron03,Shen03,Horowitz04,Burrows06,Margueron06,Gogelein07,Grygorov10,Roberts12,Horowitz12,Martinez-Pinedo12,Shen14,Rrapaj15,Roberts17,Torres-Patino19,Bauer20,Oertel20} are fundamental to understand the physics of supernova explosions \cite{Bethe90,Janka96,Burrows00} and the early evolution of their compact stellar
remnants \cite{Burrows86,Janka95}. The gravitational collapse of massive stars at the end of their thermonuclear burning leads, by means of electron capture processes, to the production of a large number of neutrinos which store and release most of the initial gravitational binding energy. In the early stages following the formation of neutron stars neutrinos are trapped because, above a critical value of the density, their mean free path $\lambda$ decreases and becomes much smaller than the stellar radius. This trapping has a strong influence on the composition and on the overall stiffness of the equation of state (EoS) of neutron stars \cite{Bombaci96,Prakash97,Vidana03,Burgio11}, being in general the physical conditions of hot and lepton-rich neutron stars substantially different from those of cold and deleptonized ones. In particular, neutrino trapping keeps the concentration of electrons so high that matter is more proton-rich in comparison with the case in which neutrinos have diffused out. Neutrinos are also very important to understand the merger  \cite{Perego14,Martin15,Frensel17,Cusinato22,Radice22} and the cooling  \cite{Shibanov96,Yakovlev01,Yakovlev04} of neutron stars. Cooling is driven first by neutrino emission mechanisms such as direct and modified URCA processes, bremsstrahlung, and Cooper pair formation, the latter operating only when the temperature of the star drops below the critical temperature for neutron superfluidity or proton superconductivity. Numerical simulations of supernova explosions and neutron star mergers, as well as, cooling calculations require not only the knowledge of the hot dense matter EoS but also a reliable description of neutrino transport. Whereas the most detailed transport codes, which solve the full Boltzmann transport equation, require the knowledge of neutrino differential cross sections, simpler transport codes only need angle averaged and/or energy averaged neutrino opacities, usually expressed in the form of neutrino mean free paths. Important sources of neutrino opacities are neutrino-baryon scattering and neutrino-baryon absorption reactions mediated, respectively, by the neutral current and the charge current of the electroweak interaction. 

In this work we consider the propagation of neutrinos in hot neutron matter and, therefore, we take into account only the contribution of the scattering of neutrinos off neutrons in the evaluation of the neutrino mean free path. The extension of the present study to the case of hot asymmetric nuclear matter, more relevant to describe the physical conditions of newly born neutron stars, will be considered in the near future \cite{Vidana22}. Here, in particular, we study the effect of chiral nuclear forces on the neutrino mean free path, being our main focus the analysis of the convergence of the dynamical structure factor and the neutrino mean free path with the order of the power counting of the chiral forces when going from the leading order (LO) to the next-to-next-to-next-to leading-order (N$^3$LO),  as well as the role played by the cut-off of these forces in the determination of these quantities. A microscopic framework based on an extension to finite temperature of the Brueckner--Hartree--Fock (BHF) approximation of the Brueckner--Bethe--Goldstone (BBG) theory is employed to descrive in a consistent way both the EoS of pure neutron matter and the dynamical structure factor. Whereas, in the last years, nuclear forces derived within the framework of chiral effective field theory ($\chi$EFT) have been largely used to determine the nuclear EoS within different many-body techniques \cite{Tews13,Carbone13,Carbone14,Hebeler15,Tews16,Drischler16,Holt17,Bombaci18,Drischler19,Piraulli19,Logoteta20,Drischler21,Sammarruca21,Keller21,Lovato22}, less attention has been paid to
the role played by these forces in calculations of neutrino processes in dense matter (see {\it e.g.,} 
Refs.\ \cite{Bacca09,Bacca12,Guo19,Riz20}).

The manuscript is organized in the following way. The neutrino-neutron scattering cross section is briefly reviewed in 
Sec.\ \ref{sec:nxs}. The role of chiral forces on the dynamical structure factor and neutrino mean free path is analyzed in Sec.\ \ref{sec:dsfnmp}. Finally, a summary and the main conclusions of this work are drawn in Sec.\ \ref{sec:suc}.

\section{Neutrino-neutron cross section scattering}
\label{sec:nxs}

In this section we briefly review the expression for the neutrino-neutron cross section scattering  in hot neutron matter. Using the Fermi golden rule (see {\it e.g.} Ref.\ \cite{Joachin75}) the cross section per unit volume (or equivalently the inverse collision mean free path) for the scattering  process, which is mediated by the neutral current of the electroweak interaction, can be written as 
\begin{widetext}
\small
\begin{equation}
\frac{\sigma(E_\nu,T)}{V}=2\int\frac{d^3\vec p_{\nu'}}{(2\pi)^3}\int\frac{d^3\vec p_n}{(2\pi)^3}\int\frac{d^3\vec p_{n'}}{(2\pi)^3}(2\pi)^4\delta^4(p_\nu+p_n-p_{\nu'}-p_{n'})(1-f_{\nu'}(E_{\nu'},T))
f_n(E_n,T)(1-f_{n'}(E_{n'},T))\frac{\langle |{\cal M}|^2\rangle}{16E_1E_2E_3E_4} \ ,
\label{eq:ncx}
\end{equation}
\end{widetext}
where the invariant transition matrix $\cal M$ reads
\begin{equation}
{\cal M}=\frac{G_F}{2\sqrt{2}}(\bar \psi_{\nu'}\gamma^\mu(1-\gamma_5)\psi_\nu)(\bar \psi_{n'}\gamma_\mu(C_V-C_A\gamma_5)\psi_n) \ ,
\label{equation}
\end{equation}
with $G_F\simeq 1.436\times 10^{-49}$ erg cm$^{-1}$ being the Fermi weak coupling constant, and $C_V=-1$ and $C_A=-1.23$ the vector and axial-vector couplings. The symbol $\langle \cdot \rangle$ denotes a sum over final spins and an average over the initial ones, $p_i=(E_i,\vec p_i)$ is the four-momentum of particle $i$, and
 $f_i(E_i,T)$ is its Fermi--Dirac distribution
 \begin{equation}
f_i(E_i,T)=\frac{1}{1+\mbox{exp}[(E_i(T)-\mu_i(T))/T} \ ,
\label{eq:fd}
\end{equation}
being $E_i$ and $\mu_i $, respectively, the single-particle energy and chemical potential of the corresponding particle. The neutron  single-particle energy and chemical potential are obtained, as it is explained below, from an extension to finite temperature of the non-relativistic BHF approximation using nuclear forces derived within the framework of $\chi$EFT.

\begin{figure*}[t!]
\centering
\includegraphics[width=1.5 \columnwidth]{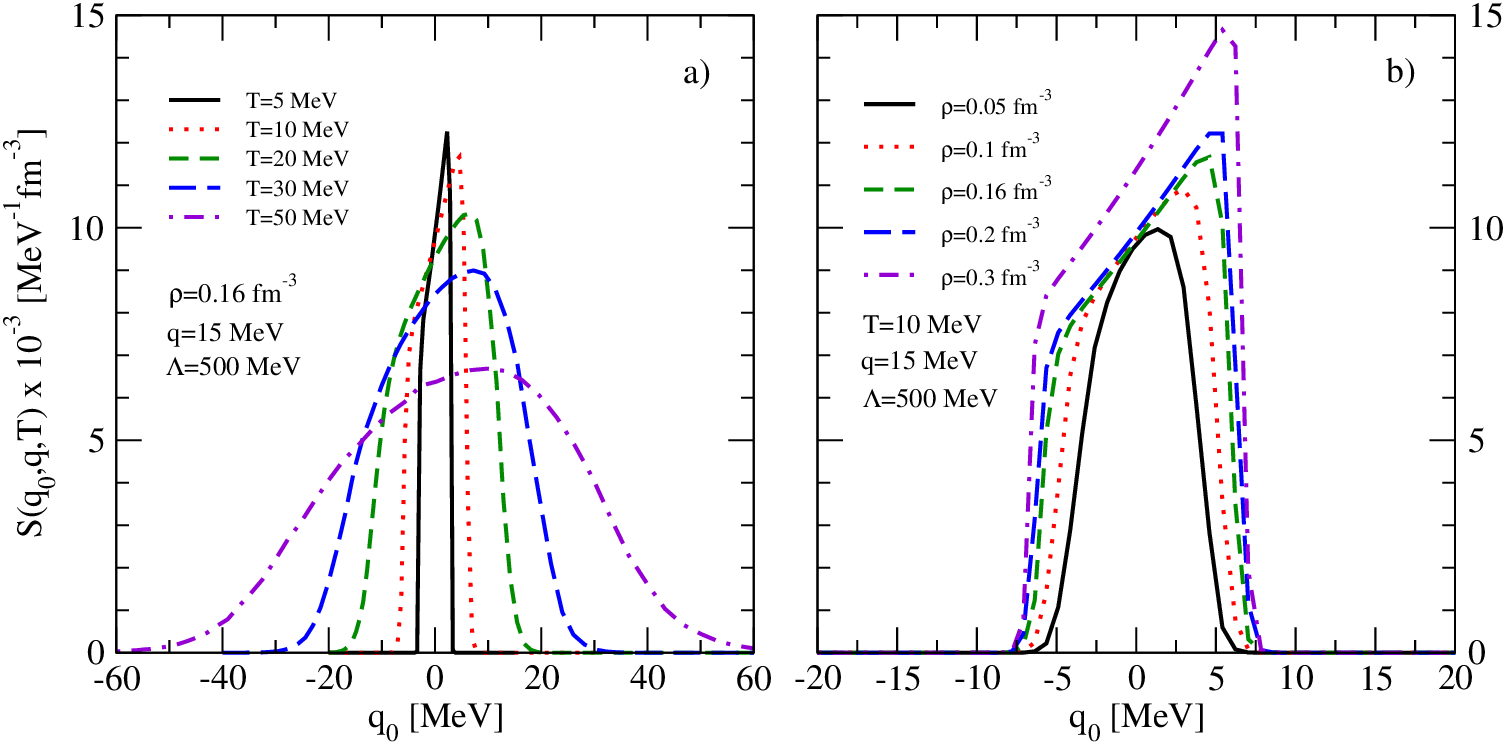}
        \caption{Energy dependence of the dynamical structure factor $S(q_0,\vec q, T)$ for several temperatures at a fixed density 
$\rho=0.16$ fm$^{-3}$  (panel $a$) and various densities at a fixed temperature $T=10$ MeV (panel $b$).  The calculation is done using the chiral potential at N$^3$LO including three-nucleon forces with a cut-off $\Lambda=500$ MeV. The transferred three-momentum is taken $|\vec q|=15$ MeV. }
\label{fig:fig1}
\end{figure*}
 
In the limit of non-relativistic neutrons and non-degenerate neutrinos, the scattering cross section simplifies and reads \cite{Iwamoto82b}
\begin{eqnarray}
\frac{\sigma(E_\nu,T)}{V}&=&\frac{G_F^2}{32\pi^2}\int d^3\vec p_{\nu'}\Big[C_V^2(1+\mbox{cos}\,\theta_{\nu\nu'})S^{(0)}(q_0,\vec q,T) \nonumber \\
&+&C_A^2(3-\mbox{cos}\,\theta_{\nu\nu'})S^{(1)}(q_0,\vec q,T)\Big] \ ,
\label{eq:ncx2}
\end{eqnarray}
where $q=p_\nu-p_{\nu'}$ is the transferred four-momentum from the neutrino to the neutron system, $\theta_{\nu\nu'}$ is the angle between the in-coming and out-going neutrino, and $S^{(S)}(q_0,\vec q, T)$ is the dynamical structure factor that describes the response of neutron matter in the spin channel $S=0,1$ to the excitations induced by neutrinos, and contains the relevant information of the nuclear medium. The vector and axial parts of the neutral current give rise, respectively, to density and spin-density fluctuations, corresponding to the $S=0$ and $S=1$ spin channels. The dynamical structure factor can be obtained from the imaginary part of the corresponding response function $\chi^{(S)}(q_0,\vec q,T)$ via the fluctuation-dissipation theorem \cite{Chomaz90}
\begin{equation}
S^{(S)}(q_0,\vec q, T)=-\frac{g}{\pi}\frac{1}{1-\mbox{exp}[-q_0/T]}\mbox{Im}\, \chi^{(S)}(q_0,\vec q,T) \ ,
\label{eq:dsf}
\end{equation}
where $g=2$ is the spin degeneracy of the neutrons and the factor $(1-\mbox{exp}[-q_0/T])^{-1}$ has the appearance of a step function except around $q_0=0$ where it diverges as $1/q_0$. However, the dynamical structure factor does not show any divergency because the imaginary part of the response function behaves as $q_0$ around $q_0=0$. The response function can be obtained from the bare one $\chi_0(q_0,\vec q, T)$ and the particle-hole residual interaction $V^{(S)}_{ph}(q_0,\vec q)$ within the random phase approximation (RPA) by solving the integral equation
\begin{equation}
\chi^{(S)}(q_0,\vec q, T)=\frac{\chi_0(q_0,\vec q,T)}{1-V^{(S)}_{ph}(q_0,\vec q)\chi_0(q_0,\vec q, T)} \ .
\label{eq:respf}
\end{equation}
In this work, however, we ignore the role played by the long-range correlations induced from the particle-hole residual interaction ({\it i.e.,} we take $V^{(S)}_{ph}(q_0,\vec q)=0$)
and, then, we simply assume $\chi^{(S)}(q_0,\vec q, T)=\chi_0(q_0,\vec q,T)$, independently of the  spin channel. Therefore, from now on, we will omit the spin index in the dynamical structure factor. Nonetheless, interactions among neutrons are taken into 
account at the mean field level in the calculation of $\chi_0(q_0,\vec q,T)$ 
\begin{widetext}
\begin{equation}
\chi_0(q_0,\vec q,T)=\int\frac{d^3\vec p}{(2\pi)^3}\frac{f_n(E_n(\vec p,T),T)-f_n(E_n(\vec p+\vec q,T),T)}{q_0-E_n(\vec p+\vec q,T)+E_n(\vec q, T)+i\eta} \ .
\label{eq:chi0}
\end{equation}
\end{widetext}
The neutron single-particle energy and chemical potential needed to calculate $\chi_0(q_0,\vec q,T)$ are obtained, as mentioned before, from an extension to finite temperature of the non-relativistic BHF approximation with chiral forces. This extension basically consists of replacing the zero temperature neutron occupation number
\begin{equation}
n(\vec k)=\left\{ \begin{array}{ll}
1, & \mbox{if $|\vec k|\leq k_F$} \\ \\
0, & \mbox{otherwise}
\end{array}
\right. 
\label{eq:occn}
\end{equation}
by the corresponding Fermi--Dirac momentum distribution when calculating the BHF single-particle energy
\begin{equation}
E_n(\vec k,T)=\frac{\hbar^2k^2}{2m}+\mbox{Re\,} U_n(\vec k,T) \ .
\label{eq:spe}
\end{equation}
Here $U_n(\vec k,T)$ is the neutron single-particle potential given at finite temperature by
\begin{eqnarray}
U_n(\vec k,T)=\sum_{\vec k'}f_n(E_n(\vec k',T),T) \phantom{aaaaaaa} \nonumber \\
\times \langle \vec k \vec k' |G(E_n(\vec k,T)+E_n(\vec k',T),T)| \vec k \vec k' \rangle_{\cal A} \ , \phantom{aaaa} 
\label{eq:spp}
\end{eqnarray}
where the $G$-matrix, describing the effective interaction between two neutrons in the presence of a surrounding medium, is obtained by solving the Bethe--Goldstone equation which at finite temperature reads schematically
\begin{eqnarray}
 G(\omega,T) = V \phantom{aaaaaaaaaaaaaaaaaaa} \nonumber \\
+\sum_{ij} V
\frac{ (1-f_n(E_n(\vec k_i,T),T)(1-f_n(E_n(\vec k_j,T),T )}
{\omega-E_n(\vec k_i,T)-E_n(\vec k_j,T) +i\eta} 
G(\omega,T)\ . \nonumber \\
\label{eq:bbg}
\end{eqnarray}
Here $V$ is the bare nucleon-nucleon interaction of which we give some details at the end of this section, and $\omega$ is the sum of the non-relativistic single-particle energies of the interacting neutrons.

We note that the self-consistent solution of Eqs.\ (\ref{eq:spe})-(\ref{eq:bbg}) requires to extract the neutron chemical potential at each step of the iterative process from the normalization condition
\begin{equation}
\rho=\sum_{\vec k}f_n(E_n(\vec k,T),T) \ ,
\label{eq:norm}
\end{equation}
with $\rho$ being the neutron density. 

Before finishing this section we would like to say a few words on the nuclear forces employed in this work. As already said, we make use of nuclear forces derived  within the framework of $\chi$EFT. In particular, we use the chiral nuclear force of Entem and Machleidt \cite{Entem17} up to N$^3$LO and consider three values of the cut-off $\Lambda$, 450, 500 and 550 MeV. The contribution from three-nucleon forces to the neutron single-particle potentials and consequently to the dynamical structure factor and the neutrino mean free path, which appear at N$^2$LO and higher others \cite{Epelbaum02,Bernard08}, is introduced in our BHF calculation by averaging over the coordinates of one of the neutrons. This leads to an effective density dependent two-body force which is added to the two-body one before solving the Bethe--Goldstone equation. The interested reader can find explicit expressions for this effective density dependent two-body force, {\it e.g.,} in Refs.\ \cite{Holt09,Kaiser19,Kaiser20}. 

In the next section we analyze the role of the cut-off dependence of the forces on the determination of the dynamical structure factor and the neutrino mean free path as well as the convergence of these quantities when considering different orders in the power counting of the chiral nuclear forces from LO up to N$^3$LO.
  

\section{Dynamical structure factor and neutrino mean free path}
\label{sec:dsfnmp}
\begin{figure}[t!]
\centering
\includegraphics[width=1.0 \columnwidth]{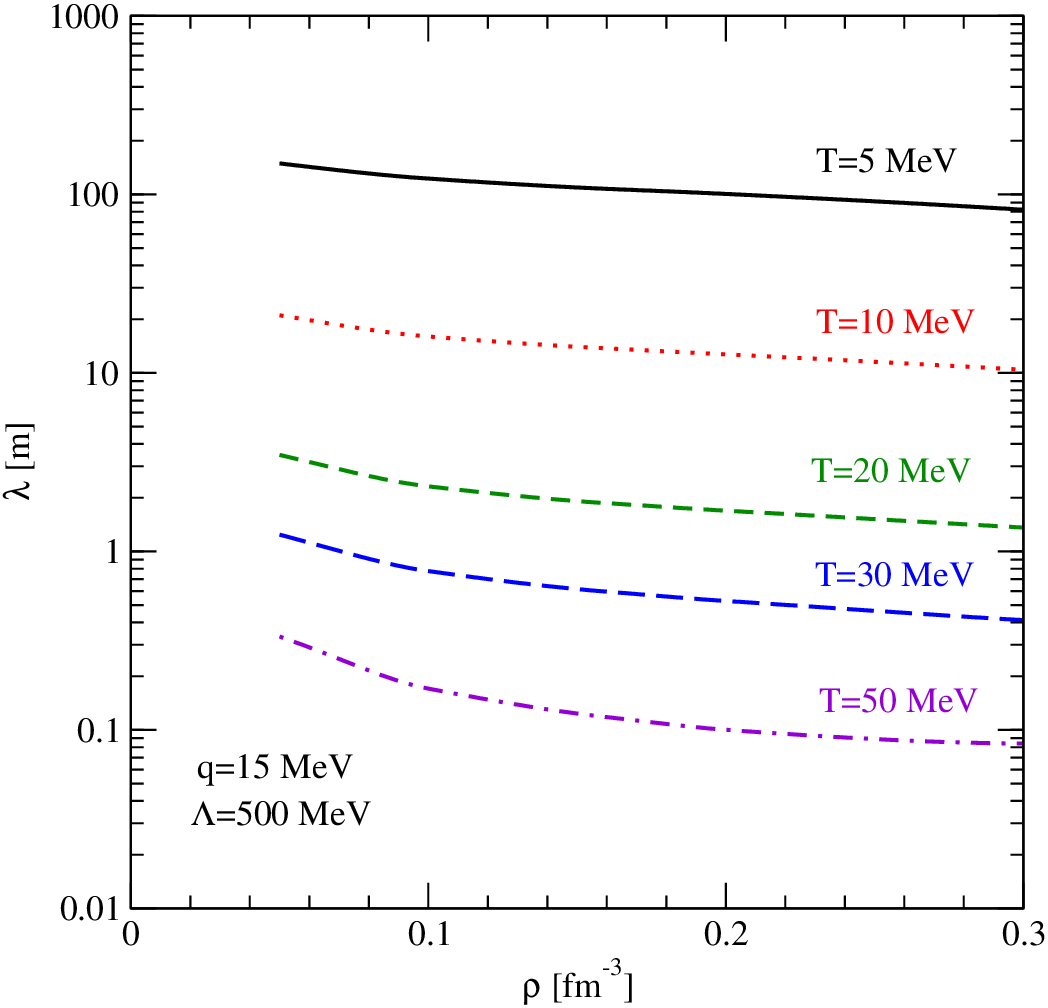}
        \caption{Neutrino mean free path as a function of the baryon number density for temperatures in the range from 5 to 50 MeV. The calculation is done using the chiral potential at N$^3$LO including three-nucleon forces with a cut-off $\Lambda=500$ MeV. The transferred three-momentum is taken $|\vec q|=15$ MeV. The energy of the neutrino is assumed to be $E_\nu=3T$.}
\label{fig:fig2}
\end{figure}

We start this section by showing in Fig.\ \ref{fig:fig1} the dynamical structure factor $S(q_0,\vec q, T)$ as a function of the transferred energy $q_0$ for several temperatures at a fixed value of the density $\rho=0.16$ fm$^{-3}$ (panel $a$) and various densities for $T=10$ MeV (panel $b$).  The calculation is done using the chiral potential at N$^3$LO including two- and three-nucleon forces with a cut-off $\Lambda=500$ MeV. The three-momentum transferred is fixed to the value $|\vec q|=15$ MeV. In this figure and all the rest, the energy of the neutrino is assumed to be $E_\nu=3T$. As it can be seen in the figure, an increase of the temperature or the density leads to a much broader dynamical structure factor with a larger area under it. The reason is simply due to the fact that the phase space of the integral in Eq.\ (\ref{eq:chi0}) increases with the temperature and the density. Consequently, an increase of the temperature or the density will give rise to a larger cross section and, therefore, to a smaller neutrino mean free path $\lambda$ when integrating Eq.\ (\ref{eq:ncx2}). This is seen in Fig.\ \ref{fig:fig2}, where  $\lambda$ is shown as a function of the baryon number density for several temperatures. We notice that the range of densities under consideration is such that the Fermi momentum is keep smaller the cut-off in all the cases. 
Note, in particular, that the neutrino mean free path varies dramatically with the temperature, decreasing about two or three orders of magnitude when increasing the temperature from 5 to 50 MeV. Since the typical radius of a neutron star is of the order of $10-12$ km,  one can easily conclude from these results that a neutrino would unlikely interact with matter at low temperatures. This conclusion is similar to those already derived by other authors. Our interest here, as we have said, is to analyze the convergence of the dynamical structure factor and the neutrino mean free path with the order of the power counting of the chiral forces, as well as their dependence on the cut-off employed. 

\begin{figure}[t!]
\centering
\includegraphics[width=1.0 \columnwidth]{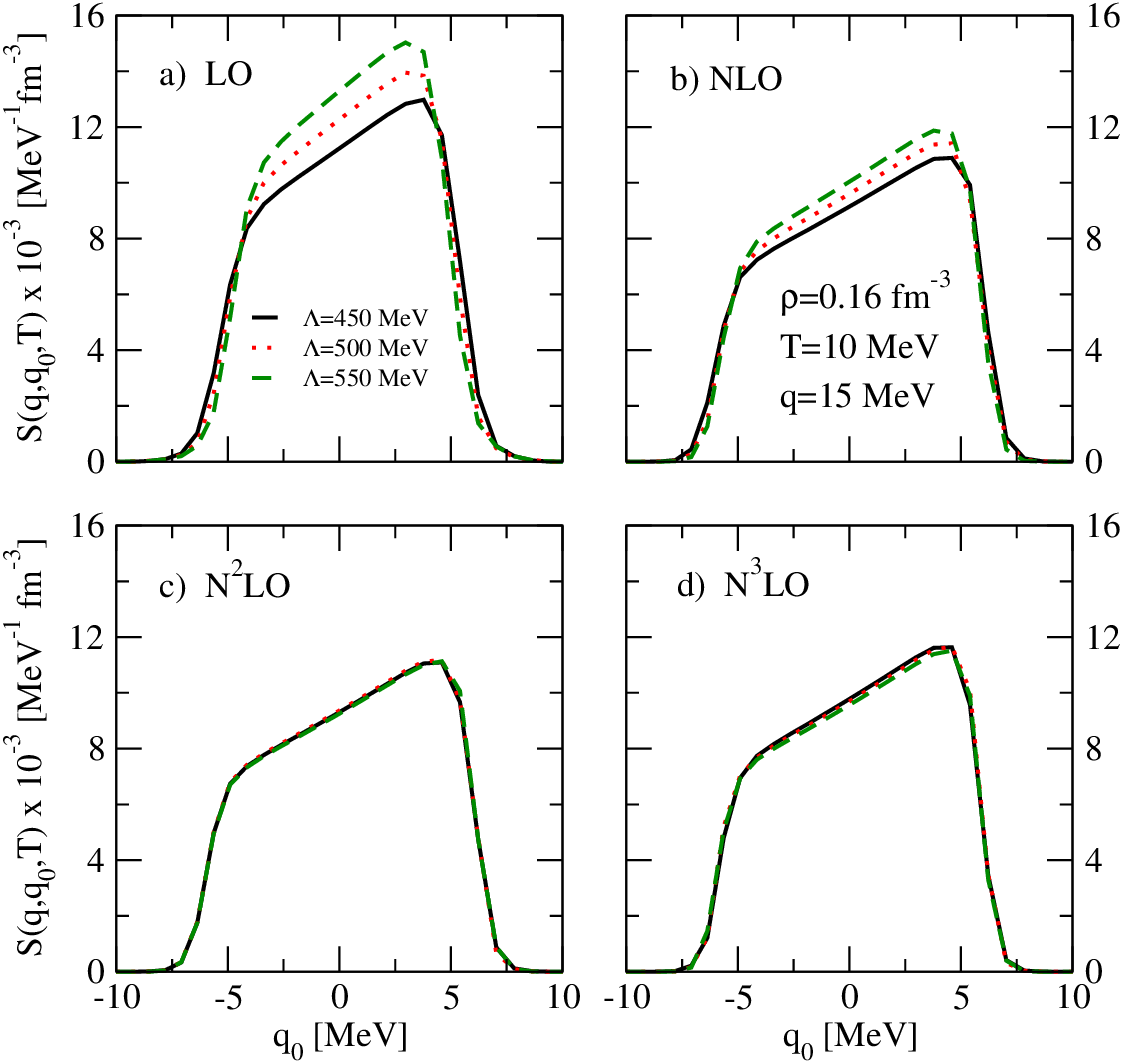}
        \caption{Energy dependence of the dynamical structure factor $S(q_0,\vec q, T)$ for several values of the cut-off $\Lambda$ at T=10 MeV and $\rho=0.16$ fm$^{-3}$. Results using the chiral potential at different orders  are shown in the different panels. The transferred three-momentum is taken $|\vec q|=15$ MeV. The energy of the neutrino is assumed to be $E_\nu=3T$.}
\label{fig:fig3}
\end{figure}

\begin{figure}[t!]
\centering
\includegraphics[width=1.0 \columnwidth]{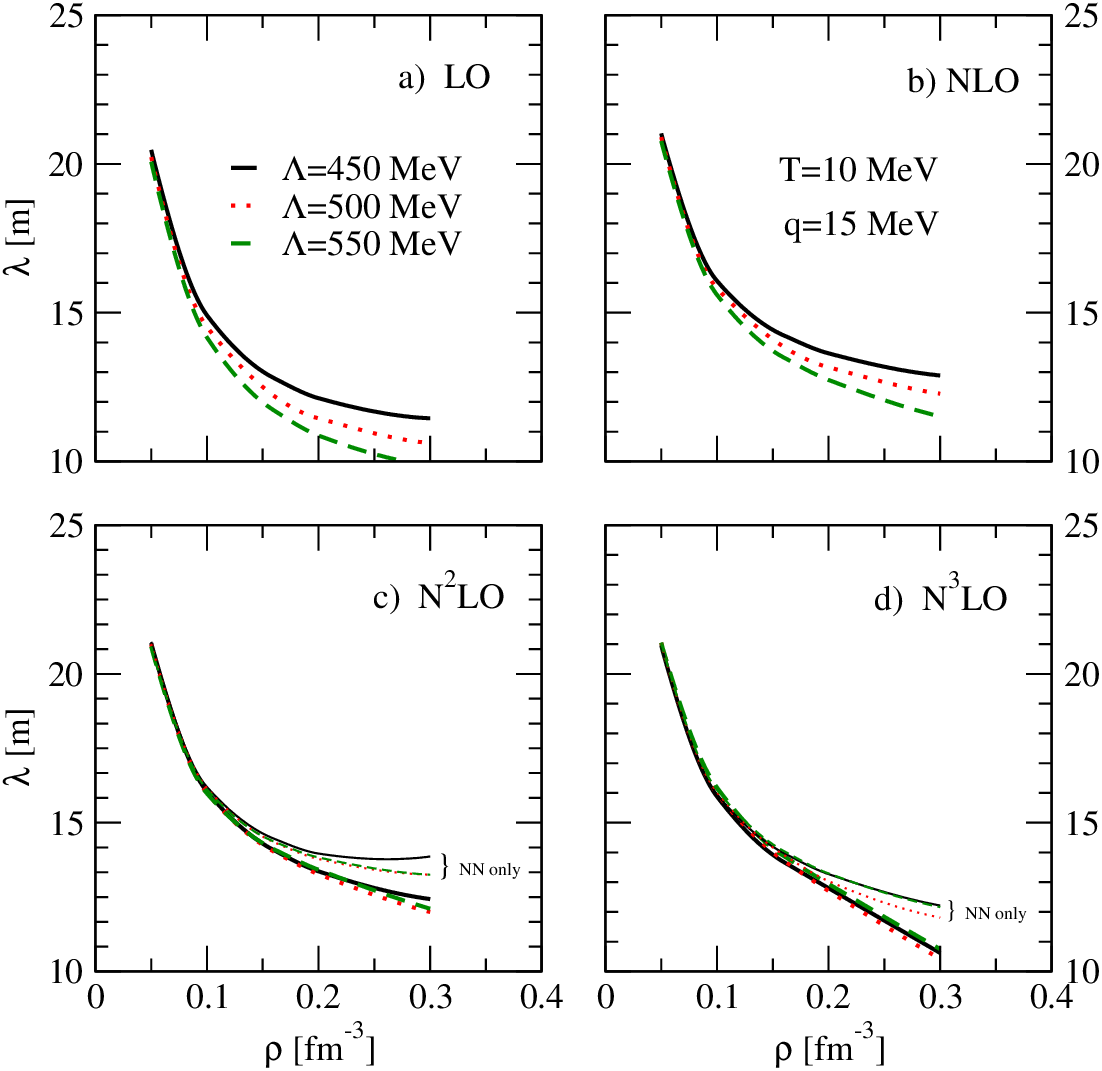}
        \caption{Density dependence of the neutrino mean free path for several values of the cut-off $\Lambda$ at T=10 MeV. Results using the chiral potential at different orders  are shown in the different panels. The transferred three-momentum is taken $|\vec q|=15$ MeV. The energy of the neutrino is assumed to be $E_\nu=3T$. Thin lines in panels $c$ and $d$ show the result when only the contribution of two-nucleon forces is taken into account in the calculation.}
\label{fig:fig4}
\end{figure}

Let  us analyze first the effect of the cut-off  of the chiral forces on the dynamical structure factor and the neutrino mean free path. In Fig.\ \ref{fig:fig3} we show the energy dependence of the dynamical structure factor $S(q_0,\vec q, T)$ evaluated at T=10 MeV and 
$\rho=0.16$ fm$^{-3}$  using the chiral potential at  LO (panel $a$), NLO (panel $b$), N$^2$LO (panel $c$) and N$^3$LO (panel $d$) for three values of the cut-off $\Lambda=450, 500$ and $500$ MeV.  The density dependence of the neutrino mean free path for the same values of the cut-off and temperature is reported in Fig.\ \ref{fig:fig4}.  Note that whereas  both $S(q_0,\vec q, T)$ and $\lambda$ show a dependence on the cut-off $\Lambda$ when the chiral potential is considered only at LO and NLO, this dependence is strongly reduced when higher order contributions to the nuclear force are taken into account, being almost negligible at N$^3$LO. This reduction of the cut-off dependence is due to the effect of three-nucleon forces that starts to contribute at N$^2$LO. This result is in agreement with the strong reduction of the cut-off dependence found for the energy per particle of neutron matter when including the contributions of three-body potentials (see, {\it e.g.,} Ref.\ \cite{Coraggio13}). To illustrate it, in panels $c$ and $d$ of Fig.\ \ref{fig:fig4} we also show (see thin lines) the results when only the contribution of two-body forces is taken into account. As it can be seen, when the contribution of three-nucleon forces is ignored, the neutrino mean free path shows a considerable dependence on the cut-off still at N$^2$LO and N$^3$LO. We should note that to restore properly the cut-off independence of the neutron matter EoS and the neutrino mean free path is crucial to treat chiral two- and three-nucleon forces consistently, that is when the same parameters are used for the same vertices that occur in all diagrams involved, as it is the case of the chiral forces employed in the present work.

\begin{figure*}[t!]
\centering
\includegraphics[width=1.5 \columnwidth]{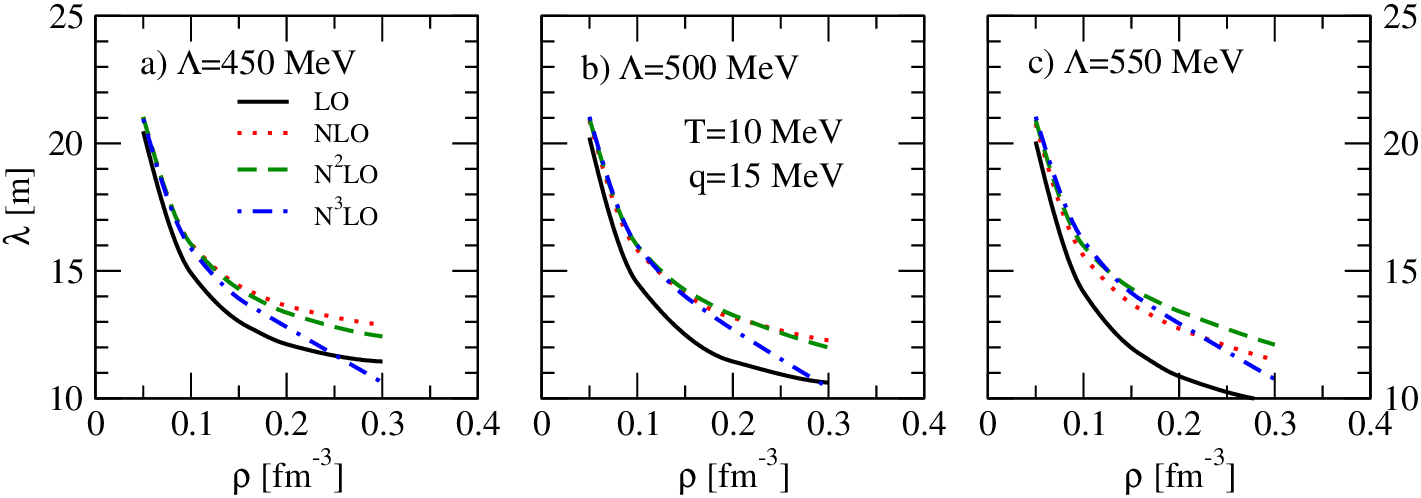}
        \caption{Dependence of the neutrino mean free path on the order of power counting of the chiral forces for the three values of the cut-off considered in all the range of densities explored at T=10 MeV for a transferred three-momentum $|\vec q|=15$ MeV. The energy of the neutrino is assumed to be $E_\nu=3T$. }
\label{fig:fig5}
\end{figure*}

Finally, we show in Fig.\ \ref{fig:fig5} the dependence of the neutrino mean free path on the order of power counting of the chiral forces for the three values of the cut-off considered in all the range of densities explored at T=10 MeV for a transferred three-momentum 
$|\vec q|=15$ MeV. As it is seen, the neutrino mean free path converges up to densities slightly below $\sim 0.15$ fm$^{-3}$ when increasing the order of the chiral power counting, being the convergence better for the two larger values of the cut-off, $\Lambda=500$ MeV and $\Lambda=550$ MeV. No signal of convergence, however, seems to exist for densities above this value. 
The resulting lack of a convergence pattern at densities larger than $\sim 0.15$ fm$^{-3}$ gives, nonetheless, an estimation of the theoretical uncertainties associated with our order-by-order nuclear many-body calculation of the dynamical structure factor and the neutrino mean free path in hot neutron matter with chiral nuclear forces. As we showed before, the results at N$^2$LO and N$^3$LO are quite independent of the value of the cut-off $\Lambda$. Therefore, the variation obtained by changing  $\Lambda$ does not seems to provide a reliable representation of the uncertainty at a given order, being a better way to estimate such uncertainty to consider instead the difference between the predictions at two consecutive orders. The difference between results for the neutrino mean free path obtained using chiral nuclear forces at N$^2$LO and N$^3$LO at T=10 MeV for a transferred three-momentum $|\vec q|=15$ MeV is shown in Tab.\ \ref{tab:tab1} for the three values of the cut-off considered and three representative densities, $\rho=0.1, 0.2$ and $0.3$ fm$^{-3}$. As it is seen, in general this difference is slightly smaller when a larger value of the cut-off is used, and it varies from about a few centimeters up to a bit less than $2$ meters in all the range of densities considered. 

\begin{table}[t]
\begin{center}
\begin{tabular}{c|ccc}
\hline
\hline
Density $\rho$ [fm$^{-3}$] &  $\Lambda=450$ MeV & $\Lambda=500$ MeV & $\Lambda=550$ MeV \\
\hline
0.1 & 0.16 & 0.07 & 0.19 \\
0.2 & 0.57 & 0.55 & 0.48 \\
0.3 & 1.81 & 1.57 & 1.35 \\
\hline
\hline
\end{tabular}
\end{center}
\caption{Absolute value of the difference between the neutrino mean free path obtained using chiral nuclear forces at N$^2$LO and N$^3$LO at T=10 MeV for a transferred three-momentum $|\vec q|=15$ MeV. Units are given in meters.}
\label{tab:tab1}
\end{table}

\section{Summary and conclusions}
\label{sec:suc}

In this work we have analyzed the role played by chiral nuclear forces on the propagation of neutrinos in hot neutron matter. In particular, we have studied the convergence of the dynamical structure factor and the neutrino mean free path when different orders in the chiral power counting are considered in the description of neutron matter, as well as as the role of the regulator cut-off of these forces in the determination of these quantities. The dynamical structure factor and the neutrino mean free path have been obtained using neutron single-particle energies and chemical potentials determined within the microscopic BHF approximation extended to finite temperature. We have found that when the chiral force is considered at N$^2$LO the dependence of the dynamical structure factor and the neutrino mean free path on the cut-off is strongly reduced and becomes almost negligible at N$^3$LO, being the three-nucleon forces, that start to contribute at N$^2$LO, the responsible for the restoration of the cut-off independence. Finally, we have found that  the neutrino mean free path converges up to densities slightly below $\sim 0.15$ fm$^{-3}$ when increasing the order of the chiral power counting, although no signal of convergence is found for densities above this one. We have roughly estimated the uncertainty associated with our order-by-order nuclear many-body calculation of the neutrino mean free path by evaluating the difference between the results obtained at N$^2$LO and N$^3$LO, finding that it varies from about a few centimeters at low densities up to a bit less than $2$ meters at the largest one considered in this work, $0.3$ fm$^{-3}$. The extension of the present study to the case of hot asymmetric nuclear matter, more relevant to describe the physical conditions of newly born neutron stars, will be considered in the near future \cite{Vidana22}.





\end{document}